# Frequency-oriented sub-sampling by photonic Fourier transform and I/Q demodulation

Wenhui Hao,[1] Yitang Dai,[1,*] Feifei Yin,[1] Yue Zhou,[1] Jianqiang Li,[1] Jian Dai,[1] Wangzhe Li,[2] and Kun Xu[1,3]

[1]State Key Laboratory of Information Photonics & Optical Communications (Beijing University of Posts and Telecommunications), P. O. Box 55 (BUPT), Beijing, 100876, China
[2]National Key Lab of Microwave Imaging Technology, Institute of Electronics, Chinese Academy of Sciences, Beijing, 100190, China
[3]School of Science, Beijing University of Posts and Telecommunications, Beijing, 100876, China
*ytdai@bupt.edu.cn

**Abstract**

Sub-sampling can acquire directly a passband within a broad radio frequency (RF) range, avoiding down-conversion and low-phase-noise tunable local oscillation (LO). However, sub-sampling suffers from band folding and self-image interference. In this paper we propose a frequency-oriented sub-sampling to solve the two problems. With ultrashort optical pulse and a pair of chromatic dispersions, the broadband RF signal is firstly short-time Fourier-transformed to a spectrum-spread pulse. Then a time slot, corresponding to the target spectrum slice, is coherently optical-sampled with in-phase/quadrature (I/Q) demodulation. We demonstrate the novel bandpass sampling by a numerical example, which shows the desired uneven intensity response, i.e. pre-filtering. We show in theory that appropriate time-stretch capacity from dispersion can result in pre-filtering bandwidth less than sampling rate. Image rejection due to I/Q sampling is also analyzed. A proof-of-concept experiment, which is based on a time-lens sampling source and chirped fiber Bragg gratings (CFBGs), shows the center-frequency-tunable pre-filtered sub-sampling with bandwidth of 6 GHz around, as well as imaging rejection larger than 26 dB. Our technique may benefit future broadband RF receivers for frequency-agile Radar or channelization.

## 1. Introduction

Radio frequency (RF) or microwave receiver is fundamental in many fields such as wireless communication, radar, and electronic warfare. Due to the urgent requirement for bandwidth in near future, receiver has to handle frequency carrier ranging tens of GHz [1]. Consequently, the flexibility of arbitrarily extracting broadband-agile signal becomes a must, adapting to typical scenarios such as operating under electronic jamming, precisely tracking some non-cooperative target, etc. Challenges then emerge for widely-employed super-heterodyne receivers, since key characteristics, like phase noise of agile local oscillation (LO) and spurious-free dynamic range (SFDR) of broadband down-conversion link, are greatly deteriorated. Besides, bandpass filter (BPF) for image rejection has usually limited tuning speed or bandwidth, while RF filter bank suffers from large volume and loss. Several optical down-conversions have been proposed to overcome the broadband difficulty, where essentially photonic super-heterodyne is implemented instead of traditional electronic one [2, 3]. Direct digital receiving is more promising, supported by rapidly-developed broadband analog to digital conversion (ADC) and digital signal processing (DSP) [4]. Photonic technologies also boost digitalization significantly in all aspects, including sampling rate and bandwidth, timing jitter, effective number of bits (ENOB), etc [5-7]. With superb flexibility, however, the processing, storage, or direct transmission of extra-large-volume data after broadband ADC requires substantial resources. *Target-oriented sampling* is therefore very attractive, aiming to compress ADC product to a more manageable rate in analog rather than by costly post digital processing. An example is photonic time-stretched ADC, where target time slot is captured and slowed down, by a broadband chirped optical pulse and large dispersion, prior to quantization [8, 9]; with a commercial oscilloscope, sampling rate as high as $10^{12}$ per second was achieved [10]. Another example is compressive sensing, where target sparse signal is sampled with repetition rate much lower than its Nyquist bandwidth, benefiting from specially encoded pulse train [11, 12]; bandwidth compression of 23.5 was achieved while 40 separated RF tones were precisely recovered [13].

Note in most scenarios, carrier frequency is still the key characteristic to distinguish the target, so a frequency-oriented sampling is highly desired. Sub-sampling, also named as bandpass sampling, provides the simplest way for direct acquirement of bandpass signal, of which the carrier frequency could locate in hundreds of GHz with photonic assistance [14]. Mixing input with a comb of oscillations rather than single LO, sub-sampling has several advantages over traditional super-heterodyne, e.g. agile LO is avoided, and ultra-low time jitter results in excellent phase noise performance at high frequency [15]. With the development of compact femtosecond fiber laser, sampling rate can match state-of-the-art high-resolution electronic ADC (several GHz). Dynamic range has also been improved by linearization technique [16]. However, band folding is inevitable: besides the target passband, any interference within the whole sampling band is down-converted to the same 1st Nyquist zone, with uniform efficiency. Reference [17] proposed an analog pre-filtered photonic ADC by shaped sampling pulse; a microwave photonic filter was incorporated without band folding, and full digitalization was still employed. An accompanying problem is self-interference by image folding, which happens when target channel covers any frequency of $kf_{\text{rep}}/2$. Here $f_{\text{rep}}$ is the sampling rate, and $k$ is an integer. Such dead zones slice the sampling band into discontinuous passbands, which limits the achievable carrier frequency, especially when instantaneous bandwidth is large (i.e. approaching $f_{\text{rep}}/2$).

In this paper we propose and demonstrate experimentally a frequency-oriented bandpass-sampling scheme to solve the above two problems, that is, pre-filtering and image interference. Signal, within duration around sampling period, is firstly real-time Fourier transformed by ultrashort optical pulse and a pair of dispersive devices with opposite value [18, 19]. Then the spectrum, already spread in time, is sampled by ultrashort pulse. Frequency-dependent sub-sampling is achieved by tuning time delay between the transformed signal and pulse gate. The image rejection is obtained by coherent in-phase/quadrature (I/Q) sampling. I/Q demodulation was first employed in optical channelization by us, where the target channel was down-converted to zero-intermediate-frequency (IF) baseband [20]. Replacing traditional continuous-wave (CW) LO by pulse train here, the target spectrum slice is moved to the 1st Nyquist zone without image folding. Instead of the original uniform one, narrow-band frequency-oriented response of sub-sampling is reported by a numerical example as well as a proof-of-concept experiment.

## 2. Principle and numerical example

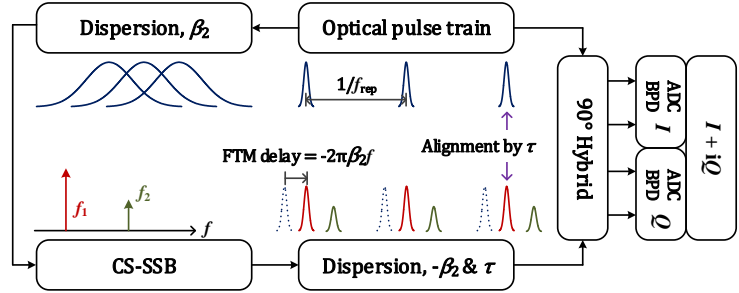

Fig. 1. Frequency-oriented bandpass sampling involving short-time Fourier transform and optically coherent I/Q sampling.

Figure 1 shows the proposed photonic bandpass sampling scheme. An optical pulse train, $s_0(t) = \sum_k q(t - k/f_{\text{rep}})$ where $q(t)$ is single pulse profile and $f_{\text{rep}}$ is repetition rate, is power divided. One part is time-stretched by a 2$^{\text{nd}}$-order chromatic dispersion device under $\beta_2$. Then the lightwave is modulated by a dual-parallel Mach-Zehnder modulator (DPMZM) which works under its carrier-suppressed, single-sideband (CS-SSB) condition. Assume the DPMZM is driven by two quadrature RF tones with frequency of $\omega_0$, and the modulated pulse passes through a second dispersion device under $-\beta_2$. The output waveform then recovers original $q(t)$ in every period, which meanwhile experiences modulation-dependent delay, $-\beta_2\omega_0$, phase shift, $-\beta_2\omega_0^2/2$, as well as the frequency shift, $\omega_0$. Mathematically,

$$s_{\text{FTM}}(t) \propto \sum_k q\left(t - \frac{k}{f_{\text{rep}}} + \beta_2\omega_0\right) e^{-i\omega_0 t - i\frac{\beta_2\omega_0^2}{2}} \tag{1}$$

where small signal approximation is assumed during CS-SSB modulation. The dispersion pair delays optical pulse according to its frequency shift, which is called "frequency to time mapping (FTM)." Assume RF signal with a complicated spectrum is input, then the variation of optical intensity with time is proportional to power spectrum of RF signal, which is the typical way for real-time Fourier transform (RTFT).

Further, pre-filtering is obtained by acquiring particular time slots of the spectrum-spread pulse train. Here we employ the other part of original pulse train as optical LO, and use a 90° hybrid coupler and two balanced photo detectors (BPDs) to implement the I/Q demodulation. $s_{\text{FTM}}(t)$ experiences a given time delay, $\tau$, beforehand. After opto-electronic conversions, two output currents, $I_{\text{Re}}$ and $I_{\text{Im}}$, are

$$I_{\substack{\text{Re}\\\text{Im}}}(t) \propto \substack{\text{Re}\\\text{Im}} \left\langle e^{-i\omega_0 t - i\frac{\beta_2\omega_0^2}{2}} \sum_k q^*\left(t - \frac{k}{f_{\text{rep}}}\right) q\left(t - \frac{k}{f_{\text{rep}}} + \beta_2\omega_0 - \tau\right) \right\rangle \tag{2}$$

Note we assume pulse duration is much less than $1/f_{\text{rep}}$, so LO pulse mixes only with its overlapped spectrum-spread pulse. Both outputs are low-pass filtered, digitalized, and numerically combined into complex quantity. Since ADC is synchronized by LO pulse train, it can be modeled by integration of each period in Eq. (2). As a result, discrete sample value after I/Q demodulation is

$$I_k \propto \left\langle e^{-i\omega_0 t} \right\rangle_{t=k/f_{\text{rep}}} \cdot \frac{R_q(\beta_2\omega_0 - \tau)}{E_q} e^{-i\frac{\beta_2\omega_0^2}{2}} \tag{3}$$

where $R_q(t)$ is autocorrelation of $q(t)$, and $E_q$ is single pulse energy which is used to normalized $R_q(t)$ [note $R_q(0)/E_q = 1$]. Equation (3) assumes the duration of $R_q$ is much less than $2\pi/\omega_0$, i.e. the optical pulse bandwidth is much larger than signal carrier frequency. Such requirement can be easily achieved by current femtosecond fiber laser.

The first term in right side of Eq. (3) means input tone, $\omega_0$, is sampled, resulting in a digital harmonic with frequency of $(\omega_0/2\pi + f_{\text{rep}}/2)\% f_{\text{rep}} - f_{\text{rep}}/2$. % is modulo operation. The second term shows such bandpass sampling is driving-frequency-dependent. Assume the duration of $R_q(t)$ is $\Delta_R$, then intensity response and the corresponding bandwidth of the proposed sub-sampling are

$$\left| H(2\pi f) \right| = \left| \frac{1}{E_q} R_q(2\pi\beta_2 f - \tau) \right|, \text{ and } B_H = \Delta_R / 2\pi |\beta_2| \tag{4}$$

, respectively. Note the sampling rate already sets an inherent Nyquist bandwidth, which is $f_{\text{rep}}$ for I/Q demodulation. If no band-folding interference is wanted, $f_{\text{rep}}$ should be larger than above pre-filtering bandwidth, that is,

$$2\pi|\beta_2|B_q f_{\text{rep}} > 1 \tag{5}$$

where $B_q$ is the bandwidth of optical pulse $q(t)$, and is inversely proportional to $\Delta_R$. Note $1/f_{\text{rep}}$ is pulse train period, so Eq. (5) means the duty-cycle after time stretch by the first dispersion is larger than 1. Such condition is easy to understand: once the duty-cycle is less than 1, part of signal in time domain is discarded, and band folding occurs.

Equation (4) shows the center frequency of target passband could be simply tuned by adjusting the time delay of spectrum-spread pulse, which is $\tau/2\pi\beta_2$. Such tunability may be with much easier implementation and faster tuning speed, since no real filter is employed. Since the spectrum spreading occurs periodically, the period, $1/f_{\text{rep}}$, sets a limited unambiguous FTM bandwidth; that is, once the modulation-induced time delay is larger than $1/f_{\text{rep}}$, the corresponding slice is mapped to the neighbor period and is mistaken for another frequency. The unambiguous bandwidth is then

$$B_f = \frac{1}{2\pi|\beta_2|f_{\text{rep}}} \tag{6}$$

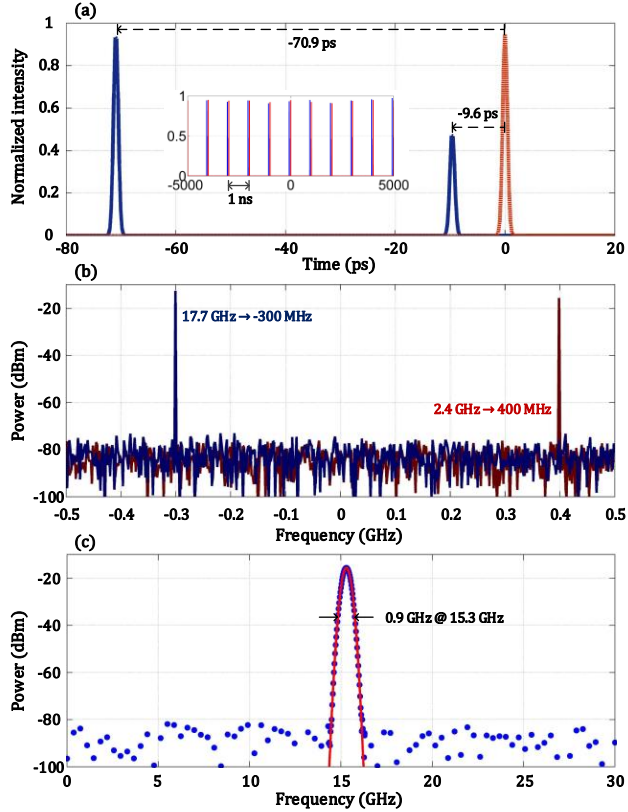

Fig. 2. A numerical example. (a) Spectrum-spread pulse train after FTM. (b) Spectrums of the 1st Nyquist zone after I/Q demodulation. (c) The frequency-oriented intensity response of bandpass sampling.

Though CS-SSB modulation is used here, regular modulation like carrier-suppressed, double-sideband (CS-DSB) by MZM also works. Under CS-DSB, two sidebands correspond to different optical frequency offset, which could still be distinguished by LO time gate after FTM. The use of SSB modulation, suppressing one sideband before mapping, enlarges the unambiguous FTM bandwidth by two. However, in practice the CS-SSB modulation may suffer from low performance under large modulation bandwidth, and regular MZM could be employed, compromising on $B_f$.

As comparison, typical sub-sampling product, under intensity-modulation direction-detection (IM-DD) scheme and small-signal sinusoidal driving, is $I_k \propto \langle \cos \omega_0 t \rangle_{t=k/f_{\text{rep}}}$. Our improved sub-sampling, besides pre-filtering, suppresses the image tone, which benefits from I/Q demodulation. Without such implementation, Eq. (2) shows two pre-filtered sub-sampling with self-image interference. In addition, Eq. (3) shows a quadratic phase response, which may be ignored if pre-filtering bandwidth, $B_H$, is small. It could also be precisely compensated digitally.

The above theory is illustrated by following numerical example. Setup follows Fig. 1. The repetition rate, average power, and pulse duration of optical pulse train are 1 GHz, 20 dBm, and 1 ps, respectively. Gaussian profile of each pulse and relative intensity noise (RIN) of -160 dBc/Hz are assumed. The followed power split ratio is 1:9, and 10% is for LO. Dispersion for stretching and compressing are -500 and 500 ps/nm, respectively. Both have loss of 5 dB. Half-wave voltage of each MZM in the DPMZM is 5 V, and the intrinsic loss is 8 dB. An Erbium-doped fiber amplifiers (EDFA) is used to boost optical power after CS-SSB modulation, with gain of 20 dB and noise figure of 5 dB. Since optical power after carrier-suppression modulation is usually low, power saturation is then ignored. Insertion loss of tunable delay is 2 dB. Each detector responsivity of BPDs is 1 A/W. We assume ADCs have high enough ENOB. Resolution bandwidth (RBW) during simulation is 1.95 MHz.

The FTM-induced pre-filtering is illustrated by inputting two RF tones, 2.4 and 17.7 GHz, and powers are 0 and 3 dBm, respectively. Figure 2(a) shows the spectrum-spread pulse train after dispersion pair. FTM results two separated pulses corresponding to input tones, of which the time delays are -9.6 and -70.9 ps, respectively, offset from the un-delayed LO pulse. This values are consistent with theoretical prediction, i.e. $-\beta_2 \omega_0$. By aligning LO pulse train with either FTM output, the corresponding tone is then down-converted to the 1st Nyquist zone after final digital I/Q synthesis, as shown in Fig. 2(b). The frequencies are 0.4 and -0.3 GHz, respectively. One observes only one digital tone within Nyquist zone, without the symmetric tone which causes self-image interference in traditional sub-sampling.

Then, under a fixed time delay, $\tau = 61.3$ ps, the driving frequency is scanned from 0 to 30 GHz while its power is kept 0 dBm. At each input, output power of the I/Q product is recorded and plotted in Fig. 2(c). The above time delay, pulse width, and dispersion pair result in pre-filtering at 15.3 GHz and 20-dB bandwidth of 0.9 GHz, less than sampling rate. Theoretical curve by Eq. (4) is also plotted. Note the unambiguous bandwidth, according to Eq. (6), is as large as 250 GHz. That is, within such a broad band only the target 15.3 GHz is extracted as shown in Fig. 2(c). The traditional band folding of sub-sampling is well suppressed. We also find a Gaussian profile of intensity response, because of the selected pulse shape. A square-like passband is expected by further pulse shaping the LO train.

## 3. Experiment

Our proposal is demonstrated by a proof-of-concept experiment. The setup is shown in Fig. 3(a). A 2.5-GHz sampling source is obtained by time lens: a CW lightwave (DFB-934 from Applied Optoelectronics) is first modulated by two phase modulators which are both driven by power-amplified 10-GHz sinusoidal wave, and then is pulse-picked by an intensity modulator which is driven by 2.5-GHz pulse train (from Agilent N4960A; pulse width is 37.5 ps). The pulse is de-chirped by matched dispersion (Finisar Waveshaper, 4000S). We use two cascaded chirped fiber Bragg gratings (CFBGs) to get sufficient time stretching, as well as another two for matched pulse compressing. The dispersion of each pair is around 1200 ps/nm within 1.5-nm passband. Time delay of LO train is controlled manually. EDFAs are used after pulse picker and after CS-SSB modulation. The measured frequency response of grating pair and spectrum of sampling pulse are shown in Fig. 3(b) and (c), respectively.

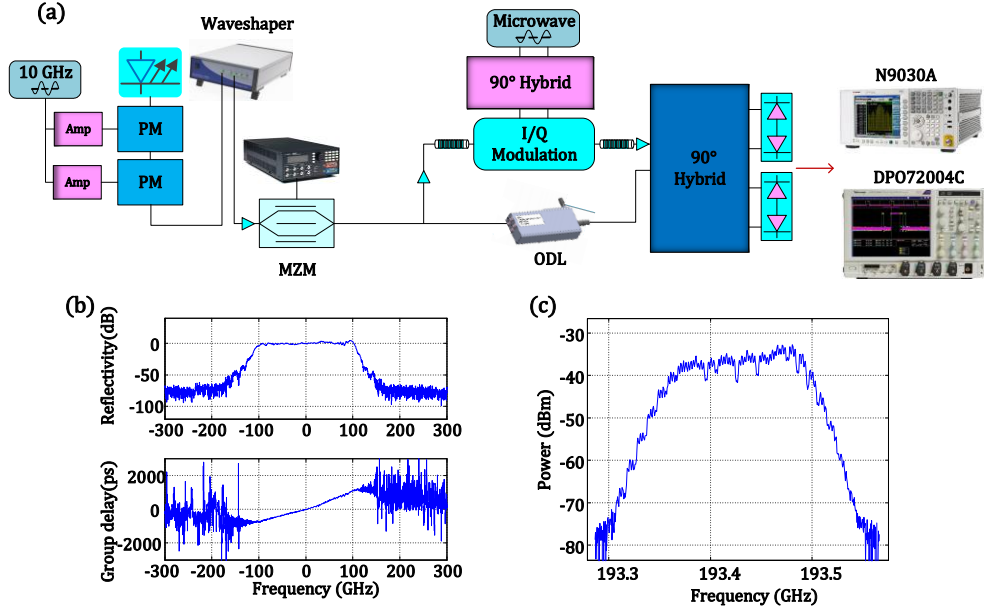

Fig. 3. (a) Experimental setup of the proposed frequency-oriented bandpass sampling. Amp: power amplifier. PM: phase modulation. MZM: Mach-Zehnder modulator. ODL: optical delay line. (b) Frequency response of cascaded CFBGs for time stretch. (c) optical spectrum of time lens source.

Firstly, the FTM is measured. When DPMZM is driven by 13.8 GHz, the delay offset of FTM output from reference LO is shown in Fig. 4(a), measured by an oscilloscope from Tektronix, DPO72004C (sampling rate is 100GS/s, without averaging). When driving frequency is tuned from 3.8 to 15.8 GHz, the delay offset is shown in Fig. 4(b), consistent with prediction (9.6 ps/GHz). Accordingly, the unambiguous bandwidth under CS-SSB modulation is 41.6 GHz.

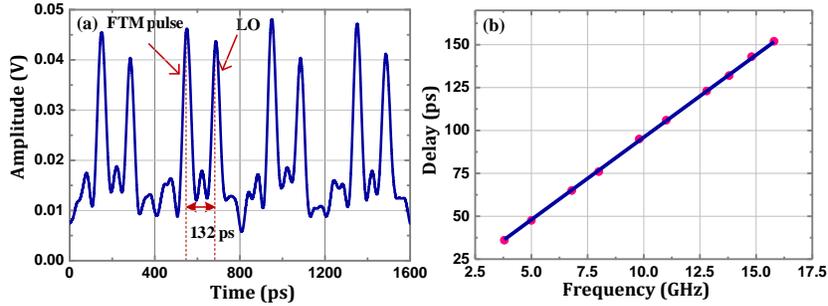

Fig. 4. The measured FTM results. (a) The delay offset between FTM output and the reference pulse under 13.8 GHz modulated frequency. (b) The measured (pink point) and theoretical relationship (blue line) of the FTM.

When RF driving is fixed while the time delay of LO train is tuned to align with FTM pulse, the target tone is successfully down-converted. Figure 5(a) and (b) shows the spectrum of I/Q product when driving frequency is 5.4 GHz and 10.8 GHz, respectively. Down-converted digital tone is at 400 MHz and 800 MHz, respectively. The image tones are suppressed by > 26 dB. In experiment, our oscilloscope cannot be synchronized by LO train, so BPD outputs are over-sampled and Fig. 5 shows the spectrum within the 1st Nyquist zone of sub-sampling to be measured.

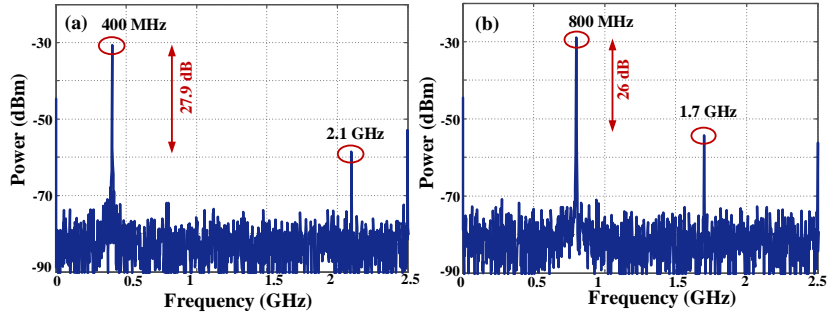

Fig. 5. The down-converted results and the image suppression effect with (a) 5.4 GHz and (b) 10.8 GHz modulated frequency, respectively.

Finally, the pre-filtering response is measured. We fix the delay of LO train and tuning the input RF frequency. The power of down-converted digital tone after I/Q synthesis is recorded in turn, and its frequency dependence is shown in Fig. 6(a) and (b), when the pre-filter is centered at 6.6 GHz and 10.8 GHz, respectively. The pre-filter has a 15-dB bandwidth around 6.2 GHz. A simulation is carried out for comparison, based on the used time-lens source and dispersion, where the bandwidth is calculated as 5 GHz. We believe the enlarged bandwidth in experiment results from the uneven intensity response of CFBG, which broadens the optical pulse after FTM. Nevertheless, the bandwidth is larger than the Nyquist bandwidth (2.5 GHz), which means band folding still occurs. Current setup is limited by our CFBG-based time-stretch capacity. Device with large "bandwidth × dispersion" is expected. In Fig. 6 we observe high side-lobes, which we believe comes from the high un-suppressed carrier as well as other sideband harmonics after DPMZM.

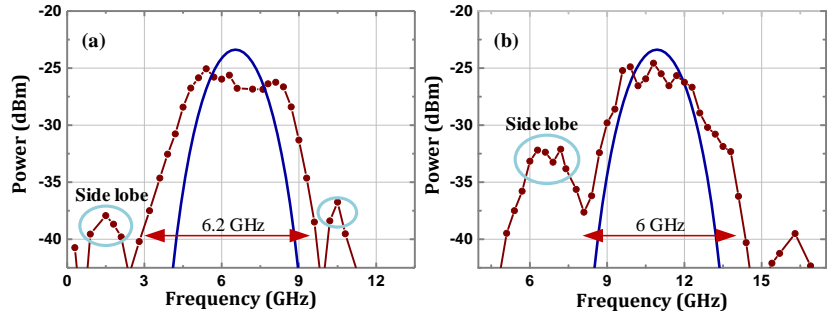

Fig. 6. The measured (red point) and the simulated (blue line) pre-filtering response when it is centered at (a) 6.6 GHz and (b) 10.8 GHz, respectively.

## 4. Summary

In conclusion, we experimentally demonstrated a frequency-oriented, photonics-assisted bandpass sampling with self-image suppression. The pre-filtering bandwidth was around 6.2 GHz, and could be tuned within unambiguous 41.6-GHz bandwidth. The self-image suppression was 26 dB. Such novel functions were obtained by short-time analog Fourier transform and coherent I/Q optical sampling. Our theory was illustrated in detail, and a numerical example was carried out to show that pre-filtering with bandwidth less than sampling rate is possible with enlarged time-stretch capacity. Advantages include easy passband tuning by time delay adjustment, and convenient promotion to multi-channel receiving, e.g. RF channelized receiver, by simple power splitting.